\documentclass[lettersize,journal]{IEEEtran}
\usepackage{amsmath,amsfonts}
\usepackage{algorithmic}
\usepackage{algorithm}
\usepackage{array}
\usepackage{caption}
\usepackage{subcaption}
\usepackage[caption=false,font=normalsize,labelfont=sf,textfont=sf]{subfig}
\usepackage{textcomp}
\usepackage{stfloats}
\usepackage{url}
\usepackage{verbatim}
\usepackage{graphicx}
\usepackage{cite}
\usepackage{makecell}
\usepackage[american,cuteinductor]{circuitikz}
\usepackage{siunitx}
\usepackage[mode=buildnew,subpreambles=true]{standalone}
\usepackage{tikz}
\renewcommand\fbox{\fcolorbox{red}{white}}
\newcommand\submittedtext{%
  \footnotesize This work has been submitted to the IEEE for possible publication. Copyright may be transferred without notice, after which this version may no longer be accessible.}

\newcommand\submittednotice{%
\begin{tikzpicture}[remember picture,overlay]
\node[anchor=south,yshift=10pt] at (current page.south) {\fbox{\parbox{\dimexpr0.65\textwidth-\fboxsep-\fboxrule\relax}{\submittedtext}}};
\end{tikzpicture}%
}
\newcommand\copyrighttext{%
  \footnotesize \textcopyright \the\year{} IEEE. Personal use of this material is permitted. Permission from IEEE must be obtained for all other uses, including reprinting/republishing this material for advertising or promotional purposes, collecting new collected works for resale or redistribution to servers or lists, or reuse of any copyrighted component of this work in other works.}

\begin{document}

\title{SAIM: Scalable Analog Ising Machine for Solving Quadratic Binary Optimization Problems}

\author{Sasan Razmkhah, Jui-Yu Huang, Mehdi Kamal, ~\IEEEmembership{Senior Member, IEEE}, Massoud Pedram,~\IEEEmembership{Fellow, IEEE}
\thanks{This work has been funded by the National Science Foundation (NSF) under the project Expedition: (Design and Integration of Superconducting Computation for Ventures beyond Exascale Realization) project with grant number 2124453. (Corresponding author: S. Razmkhah)\\
S~Razmkhah, J.-Y.~Haung, M~Kamal, and M~Pedram are with the Ming Hsieh Department of Electrical Engineering, University of Southern California, Los Angeles, California, USA (e-mail: razmkhah@usc.edu, pedram@usc.edu)}
}

\markboth{}%
{Shell \MakeLowercase{\textit{et al.}}: A Sample Article Using IEEEtran.cls for IEEE Journals}

\maketitle

\submittednotice

\begin{abstract}
This paper presents a CMOS-compatible Lechner-Hauke-Zoller (LHZ)--based analog tile structure as a fundamental unit for developing scalable analog Ising machines (IMs). In the designed LHZ tile, the voltage-controlled oscillators are employed as the physical Ising spins, while for the ancillary spins, we introduce an oscillator-based circuit to emulate the constraint needed to ensure the correct functionality of the tile. We implement the proposed LHZ tile in 12nm FinFET technology using the Cadence Virtuoso. Simulation results show the proposed tile could converge to the results in about 31~ns. Also, the designed spins could operate at approximately 13~GHz. 
\end{abstract}


\section{Introduction}
\IEEEPARstart{T}{he} increasing complexity in computing, driven by rising demands for computational power in fields like artificial intelligence (AI) \cite{desislavov2023trends}, and combinatorial optimization problems \cite{yang2022review}, necessitates the exploration of novel architectures. As conventional CMOS technology and architectures near their physical limits, surpassing computational limits by resorting to innovative hardware technologies and architectures becomes crucial.

Several large problems fall within the NP-complete (Non-deterministic Polynomial complete) problem class \cite{garey1974some}, for which solutions cannot be found in polynomial time \cite{karp1975computational}. As the size of these problems increases, computational energy demands increase greatly, necessitating heuristics \cite{kesavan2020heuristic} and approximations \cite{hochba1997approximation}. NP-complete problems can be polynomial-time reduced to one another. Therefore, if an NP-complete problem can be modeled as a physical system, we can leverage reductions from other problems to that problem and solve the resultant physical system, leading to feasible solutions \cite{lucas2014ising}.

In the Ising model, the arrangement of electron spins in a material's lattice determines its magnetic properties. In this system, increasing the temperature raises the kinetic energy of the electrons, causing their spins to oscillate freely. As the temperature decreases, the kinetic energy of the electrons becomes negligible compared to their interactions with the lattice and other electrons. Like any other physical system, this system of interacting spins tends to settle in the minimum energy level, known as the ground state \cite{vadlamani2020physics}. 

Each physical system can be described by the sum of its kinetic and potential energy, known as its Hamiltonian. The ground state of a system is the eigenvector to this Hamiltonian relation. Therefore, NP-complete problems are solved by mapping them to the system of interacting spins in a lattice. The minimum energy of the system provides the solution for the Ising model and the problem. A system of interacting oscillators that the Ising model can describe to solve its Hamiltonian equation is called an Ising machine (IM). 

It is important to recognize that the general form of the Ising model is NP-complete, indicating that finding the exact ground state (optimal solution) for complex Ising models can be computationally challenging. While Ising formulations do not solve NP-complete problems exactly, they can still be advantageous for certain applications. Ising models, viewed as physical systems of interacting spins, often enable efficient approximate solutions, making them valuable tools for addressing many NP-complete problems. In particular, IMs excel at finding near-optimal solutions to combinatorial optimization problems in a category known as quadratic unconstrained binary optimization (QUBO) \cite{lucas2014ising}. Other NP-complete problems can be addressed by mapping them to the QUBO framework. Unfortunately, the physical implementation of IMs often limits the size of QUBO problems due to the constraints imposed by available hardware graph structures.

To date, various technologies have been employed to implement IMs. These include superconductor-based IMs, which operate at 4K temperatures \cite{razmkhah2024josephson}, quantum annealers, such as D-Wave Two Quantum (DW2Q) from D-Wave company, which function at mK temperatures \cite{hamerlyquantum},  coherent IMs \cite{optical_ising1}. Cmos-compatible IMs \cite{cmos_ising1, cmos_ising2}. Among these implementations, the fabrication complexity of CMOS-compatible IMs is significantly lower due to the maturity of available CMOS technologies. However, similar to other technologies, scalability remains the primary challenge CMOS-compatible IMs face. The Ising machine requires all-to-all connectivity, which makes its implementation extremely challenging, and as a result, the proposed structures for CMOS-compatible IMs are not available. 

To address the scalability issues and mapping QUBO problem on annealers, Lechner, Hauke, and Zoller came up with a new architecture for IMs \cite{lhz}. This architecture, known as LHZ, maps the interactions of an all-to-all connected system to the local interactions of the node of a tiled system. The tile is the unit cell of the hardware, consisting of four nodes interacting with each other under a constraint known as a penalty term. Each tile interacts with its neighbors, achieving long-range interactions between nodes.

This paper will present a CMOS-compatible scalable analog Ising Machine (called SAIM), which is based on the LHZ architecture. Thus, in this work, we introduce an analog LHZ tile structure. For this structure, we provide two circuits to emulate the required spins in the LHZ tile. One is based on a modified design of a prior voltage-controlled oscillator (VCO) circuit, which will serve as the physical spins. The other is a VCO circuit structure designed to simulate the necessary ancilla spin to model the required penalty of the LHZ tile, ensuring the correct functionality of the tile. In the LHZ-based Ising machine, the circuit connections are fixed, and weights of the given problem are applied through pump signals to the physical spins. The tile is implemented on the 12nm process in Cadence Virtuoso and simulated using Cadence Spectre to verify the functionality of the proposed structure. The results show the high stability of the proposed structure.

The remainder of the paper is organized as follows. In Section II, we describe the methodology and the architecture that is used for Ising machine implementation in this work. Section III describes the analog circuit for ideal parameters and then, for example, the 12nm CMOS process. In section IV, we discuss the results from 12nm CMOS and the power needed.

\vspace{-1em}
\section{Methodology}
\subsection{Ising Machine}
An IM is a specialized annealer machine that solves combinatorial optimization problems based on the Ising model. In an IM, spins interact with each other and can be either up (1) or down (-1). The minimum energy needed by the system will be achieved under the optimal solution of the problem. The Hamiltonian equation, which gives the system's energy, is described by
\begin{equation}
\label{eq:hamiltonian}
H = - \sum_{i,j} J'_{ij} \sigma_i \sigma_j - \sum_{i} h_i \sigma_i,
\end{equation}
\noindent 
where $h_i$ represents the local field acting on spin $i$, $\sigma_i$ denotes the orientation of spin $i$, and $J_{ij}$ signifies the interaction strength between spins $i$ and $j$.

\begin{figure}[ht]
\centering
\includegraphics[width=0.8\linewidth]{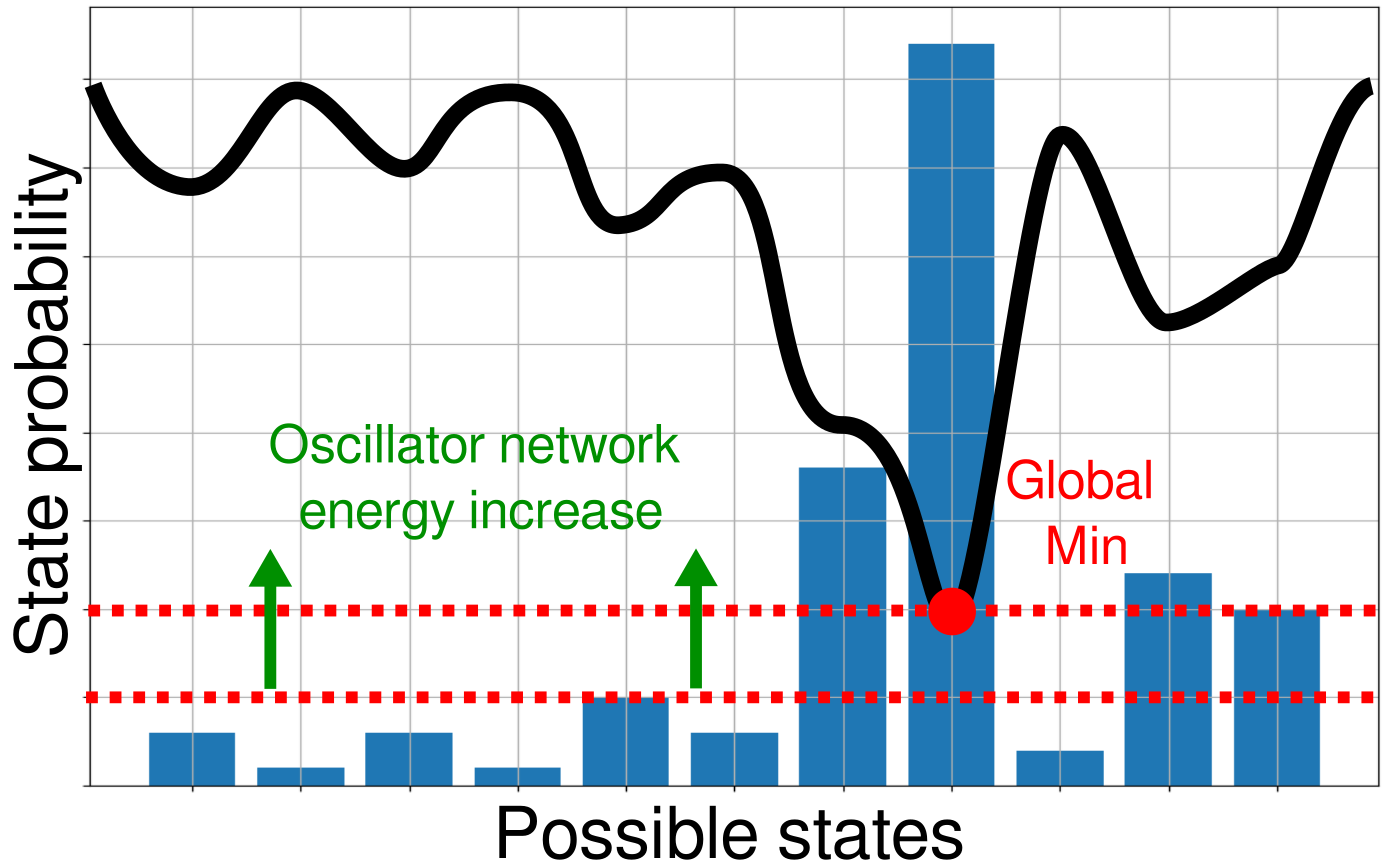}
\caption{The annealing process of a leaky oscillator network is demonstrated. If the rate of power gain does not exceed the rate of power loss for any state, none of these states can be sustained, and noise will dominate. The first stable state to emerge is the one with the smallest loss. Due to the mapping from loss to the figure-of-merit, this physical state encodes the global minimum of the problem.}
\vspace{-1em}
\label{fig:Annealing}
\end{figure}

The annealing process in the IMs varies depending on the specific structure of the IM and the technology used to fabricate it. Physical annealing involves the slow cooling down of the system, which decreases the kinetic energy of the electrons to the point that the spins are not freely oscillating and settle in a state, which translates to gradient descent in the possible state space to the ground state, as shown in Fig.~\ref{fig:Annealing}. Simulated annealing mimics this behavior by combining a greedy search method with random steps (noise) to avoid local minimus. In quantum annealers, the system is put in the ground state of an initial easy problem and then, by adiabatic annealing, moves to the hard problem's ground state \cite{vadlamani2020physics}. Here, as demonstrated in Fig.~\ref{fig:Annealing}, the leaky oscillating network will gain energy until it reaches the first stable state with the smallest loss.

\subsection{LHZ Architecture}
\label{sec:architectrure}
As stated, scalability is the fundamental challenge of the IM implementations. LHZ addresses this issue by converting all-to-all connectivity to two-by-two connectivity. To accommodate all interaction matrix elements, the system size in LHZ architecture is enlarged from N logical spins to $K = N(N-1)/2$ physical spins. 

\begin{figure}[ht]
\centering
\includegraphics[width=0.95\linewidth]{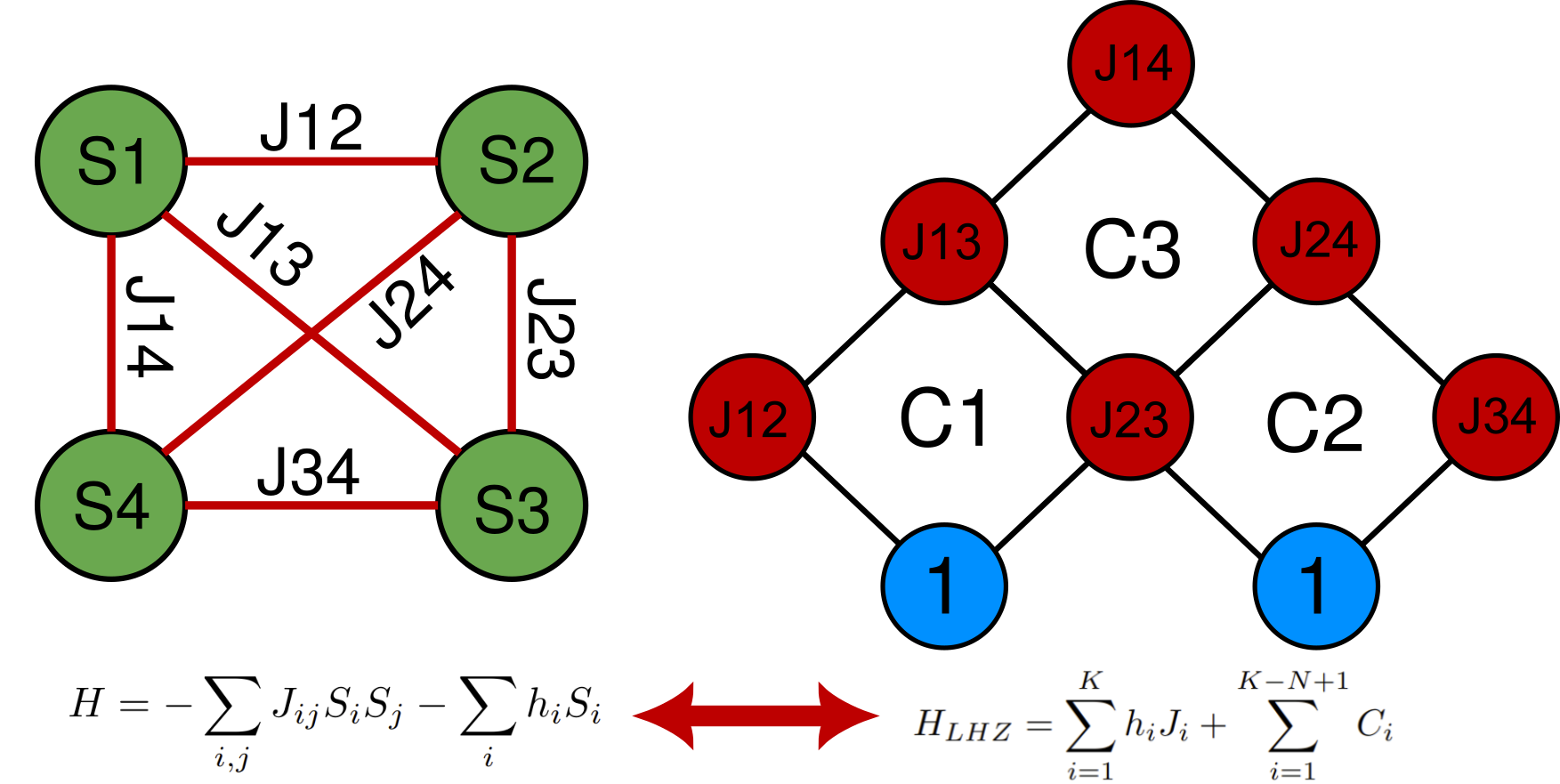}
\caption{Converting a 4-node all-to-all connected graph to the LHZ architecture. Here, the weight of the edges in the graph is mapped to the local field values of the LHZ. $J_{ij}$ demonstrates an individual node, while C1, C2, and C3 are penalty terms. The two nodes at the bottom always have a fixed value of one.}
\vspace{-1em}
\label{fig:LHZ_maxCut}
\end{figure}

In the LHZ architecture, each interaction between two logical spins is represented by a physical spin. Its local field demonstrates the interaction between two corresponding logical spins. Thanks to this structure, all-to-all interactions are converted to the local fields. Owing to this feature, the LHZ structure is scalable. Additionally, LHZ mapping has an additional constraint parameter $C$, also known as a penalty term. The Hamiltonian of the LHZ is obtained by
\begin{equation}
\begin{aligned}
H_{LHZ} = \sum_{i=1}^{K} J_i \tilde{\sigma_z}^{(i)} +  \sum_{i=1}^{K-N+1} C_i\\
C_i = C(\sum_{k=1,2,3,4} \tilde{\sigma_z}^{(i,k)}+S_z^l)^2
\end{aligned}
\label{eq:lhz_hamiltonian}
\end{equation}

\noindent where $\tilde{\sigma}$ is the physical spin, $J$ is the local field which is generated based on $J'$ in Eq.(\ref{eq:hamiltonian}), and $C$ is the constraint. The number of penalty terms ($C_i$) is $K-N$, and they are defined to guarantee that the number of positive spins in each of the tiles is always even. These constraints must be satisfied to ensure that the LHZ structure correctly models the all-to-all interactions of the IM. The penalty term is defined based on the hardware structure.
In Fig.~\ref{fig:LHZ_maxCut}, the corresponding LHZ structure of a 4-node graph (e.g., 4-node Max-Cut problem) which contains 3 LHZ tiles is shown.
Note that each weight value of the given QUBO ($w_{ij}$) is mapped to the local field of the corresponding node of the LHZ $(J_{i'}$).

\section{Circuit implementation of analog LHZ tile}
\begin{figure}[ht]
\centering
\includegraphics[width=0.7\linewidth]{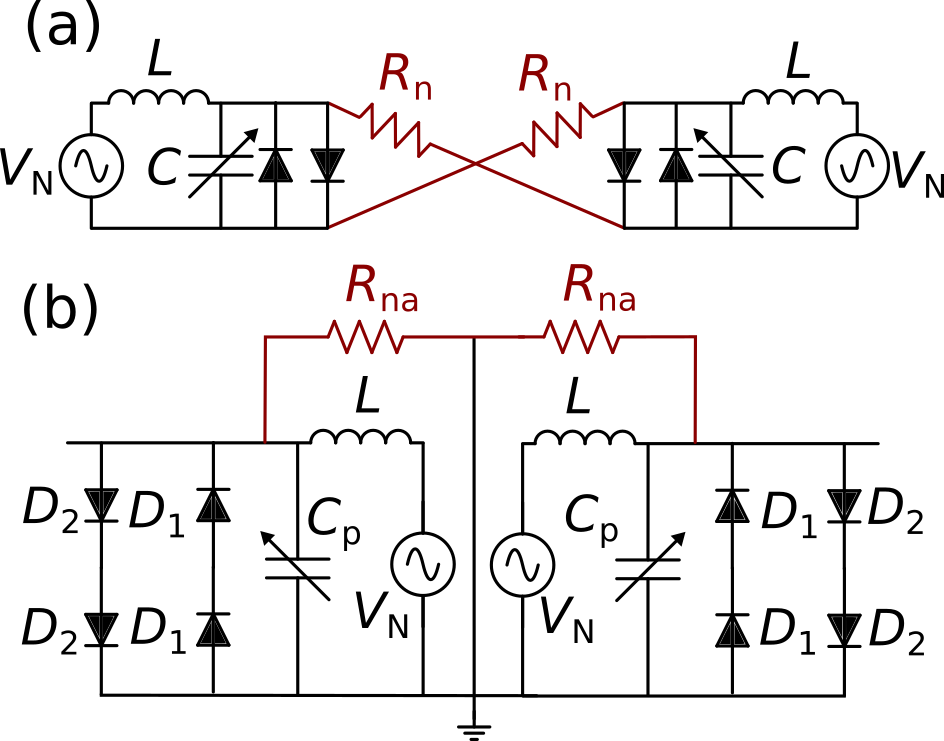}
\caption{Circuit implementation with ideal parameters, (a) parametric oscillators with anti-ferro interaction, here $L=159.15pH$, $C=159.15pF$, $R_n=200\Omega$ and diodes are 1N4148, the pump is applied to the capacitor's charge (b) Ancillary circuit is demonstrated. Here, the $R_{na}=46\Omega$, and the pump amplitude is twice the node values.}
\label{fig:Nodes}
\end{figure}

To implement the physical spins, we employ the spin structure, which has been presented in \cite{vadlamani2020physics}. The spin circuit structure is an ideal LC oscillator. Fig.~\ref{fig:Nodes}(a) demonstrates these spins. Since the interaction between the spins is fixed in the LHZ structure, we utilize fixed-value resistors in the antiferromagnetic (aF) scheme to make fixed negative interactions between nodes. In our design, the local fields are controlled by the electric charge pumped on the parametric capacitor, and they are changed based on the weights of the given QUBO problem. The capacitor's charge value is described as,
\begin{equation}
    Q = ((G_0 + dG\frac{t}{T_a})cos(\omega_p t))C_0\times V_C
\end{equation}
where $Q$ is the capacitor charge, $G_0$, and $dG$ are constants defining the pump's amplitude and its changing rate function for annealing, respectively. $T_a$ is the annealing time, $\omega_p$ is the pump frequency, $C_0$ is the capacitor value and $V_C$ is capacitor's voltage. The diodes are exploited to limit the output voltage of the LC oscillators.

To capture the LHZ conditions, we need to add the penalty function implemented via ancillary nodes. Fig.~\ref{fig:Nodes}(b) shows the proposed ancillary circuit. The ancilla nodes should be able to generate  4$\times$, 0, or -4$\times$ kinetic energy compared to the normal nodes to fulfill the LHZ condition as explained in section~\ref{sec:architectrure}. The ancilla circuit contains two LC oscillators, which are connected in an aF scheme with higher interaction than other aF connections. Here, the negative output of the ancilla is grounded, and the two ports are from the positive outputs of the LC oscillators. The charge pump amplitude described by $G_0$ and $dG$ constants in the capacitor equation of ancilla are twice their counterparts in normal nodes.

\begin{figure}[ht!]
\centering
\includegraphics[width=0.7\linewidth]{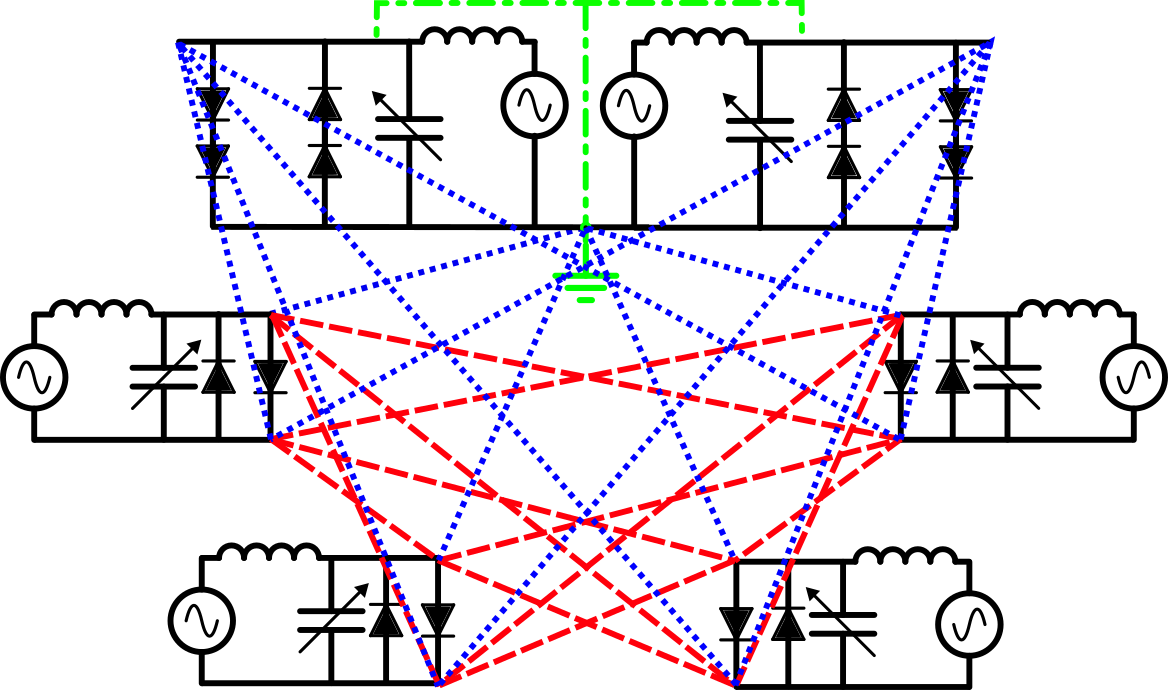}
\caption{A tile implemented with the ideal nodes. Three connections are needed to get the correct ground states expected from the LHZ tile. Here, the connection between nodes is $200\Omega$, between ancilla and nodes $250\Omega$, and inside ancilla is $46\Omega$. The tile is repeated in the LHZ IM, and all the tiles are similar.}
\vspace{-1em}
\label{fig:TileSceme}
\end{figure}

A single tile's connections are shown in Fig.~\ref{fig:TileSceme}. Here, three different resistor values for aF connections are used: between nodes, between node and ancilla, and in the ancilla circuit. These resistors determine the limits of interactions between different nodes, and they are fixed for all the problems that we map on the IM.

\begin{figure}[ht!]
\centering
\includegraphics[width=0.8\linewidth]{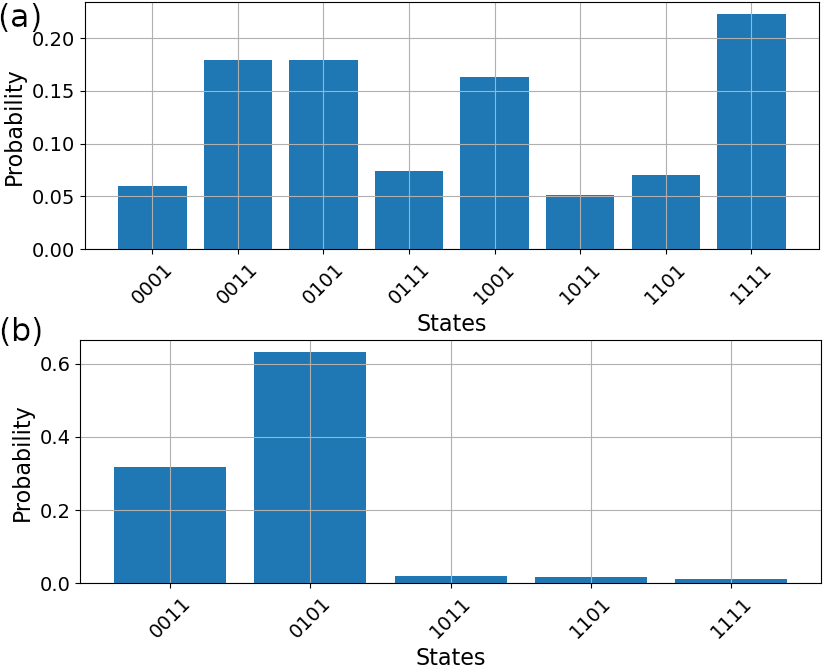}
\caption{Simulation results of 1000 runs for a single tile and 4-node LHZ network, (a) tile simulation results show the high probability of even states, (b) 4-node LHZ network simulation results for an unweighted Max-Cut network. The correct results are two ones and two zeros. By shifting the annealing time, we can get different correct answers.}
\vspace{-1em}
\label{fig:SimResult}
\end{figure}

To verify this structure, we have implemented a technology-independent single LHZ tile based on the structures and values provided in Figures \ref{fig:Nodes}(b) and \ref{fig:TileSceme}. Three tile structures are also employed to implement an Ising machine for solving unweighted Max-Cut (Fig. \ref{fig:LHZ_maxCut}).
The simulation results for 1000 runs on the noisy circuits are shown in Fig.~\ref{fig:SimResult}. The simulations were done with the NgSPICE engine. We extracted the output waveform of each node and, using FFT, calculated the phase of oscillations. Since the global phase doesn't affect the ground states, the mirror of ground states is also a valid answer. We always assume the first node is one, and based on the phase difference, we calculate the value of other nodes. Fig.~\ref{fig:SimResult} shows the simulation results for a (a) tile and a (b) 4-node network. 

\subsection{Implementation on 12~nm process} 
\begin{figure}
    \centering
    \includegraphics[width=.8\linewidth]{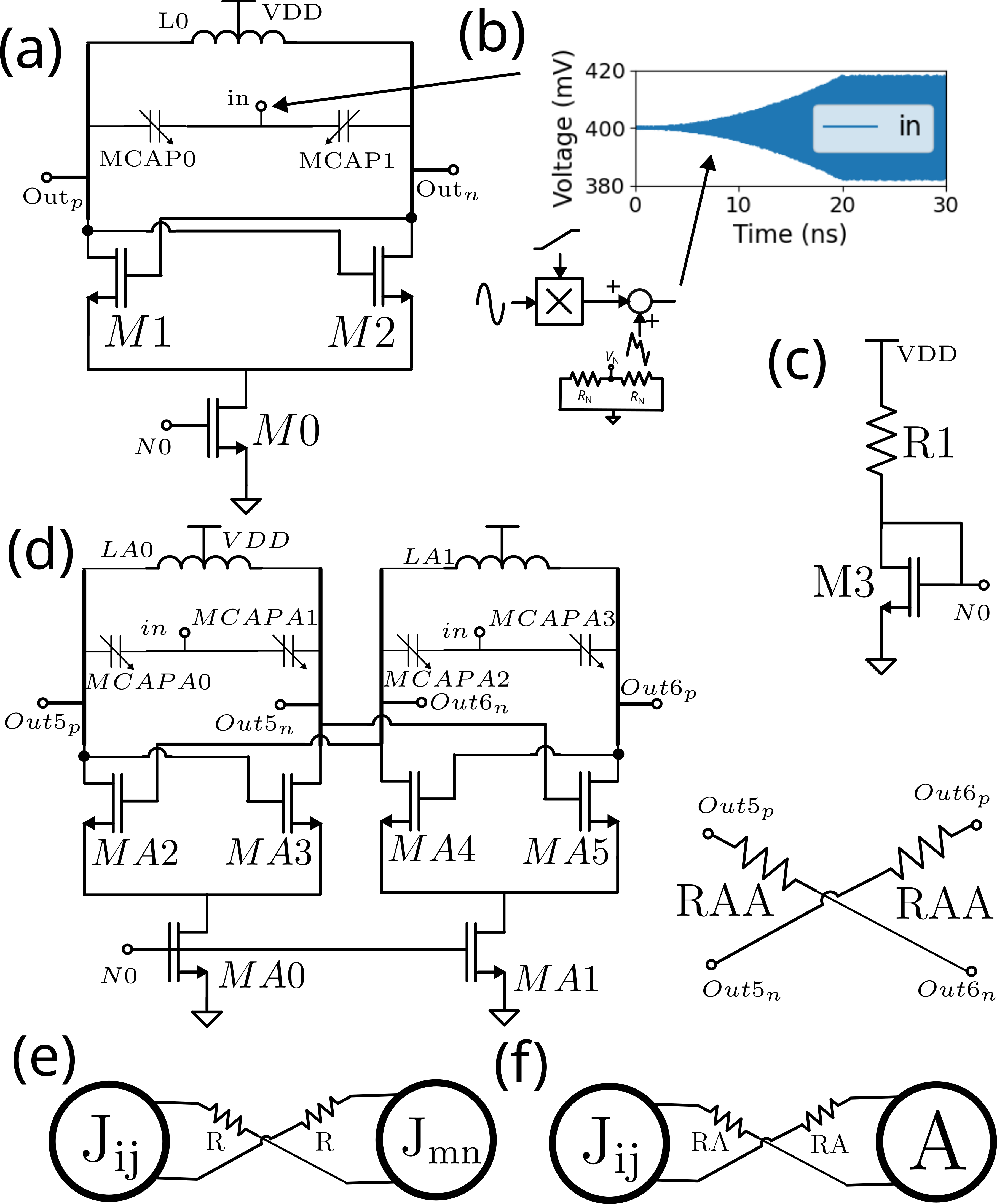}
     \caption{Parametric Oscillator and ancilla circuit design for a single tile: (a) is the schematic of a single LHZ node; Here W/L for M0, M1, M2, MCAP is $\qty{5.2}{\um} / \qty{14}{\nm}$, $\qty{2.3}{\um} / \qty{14}{\nm}$, $\qty{192}{\nm} / \qty{14}{\nm}$, $\qty{3.4}{\um} / \qty{80}{\nm}$ respectively.  (b) displays the input pump and the annealing function for each node; (c) represents the schematic of the current source from which the LHZ and ancillary nodes copy; (d) shows the ancillary node design with internal antiferromagnetic connections; (e) and (f) illustrate the connections between two LHZ nodes and between an LHZ node and an ancillary node, respectively.}
     \label{fig:SingleTileCMOS}
\end{figure}

After validating the design with ideal oscillators, the tile circuit of an LHZ network was implemented with 12~nm CMOS technology. The parametric oscillators are based on the discussed LC oscillators. The schematic and parameters of the four spins and an ancillary node of the LHZ tile are illustrated in Fig.~\ref{fig:SingleTileCMOS}. The oscillation frequency of a single oscillator in the target technology is \qty{13.37}{\GHz}. In Fig.~\ref{fig:SingleTileCMOS} (a), the \textbf{in} terminals are injected with an annealing signal with added thermal noise shown in Fig.~\ref{fig:SingleTileCMOS} (b). This pump signal increases quadratically with time until the annealing time $\Delta t$ is reached. Therefore, the pump signal is described as $V_{AC}(t)\times(\frac{t}{\Delta t})^2$ for $t \leq \Delta t$ and $V_{AC}(t)$ for $t > \Delta t$, where $V_{AC}(t)=\qty{18}{\mV}\times \sin{2\pi f_p t}$, $\Delta t = \qty{20}{\ns}$, and $f_{p}=\qty{26.74}{\GHz}$. 

The thermal noise is generated from two parallel \qty{10}{\ohm}-resistors. The expression for the noise spectral density~\cite{razavi2016analog} is $S_{v} = 4kT(R_N/2)$, where $k=\qty[per-mode=symbol]{1.38e-23}{\J\per\K}$ is the Boltzmann constant, $T$ is the temperature, and $R_N=\qty{10}{\ohm}$. At $T=\qty{300}{\K}$, the resistors generate \qty[per-mode=symbol]{8.28e-20}{\volt^2\per\Hz} of thermal noise. These $V_{AC}$ and thermal noise signals with the resistors connecting two oscillators should be optimized to ensure uniformity and a high probability of even states. The maximum oscillator amplitude is controlled by the transistors at the current source, M0 and MA0, and their channel width is controlled to ensure the LHZ conditions.

\section{Results and Discussion}
\begin{figure}[ht]
\centering
\includegraphics[width=0.9\linewidth]{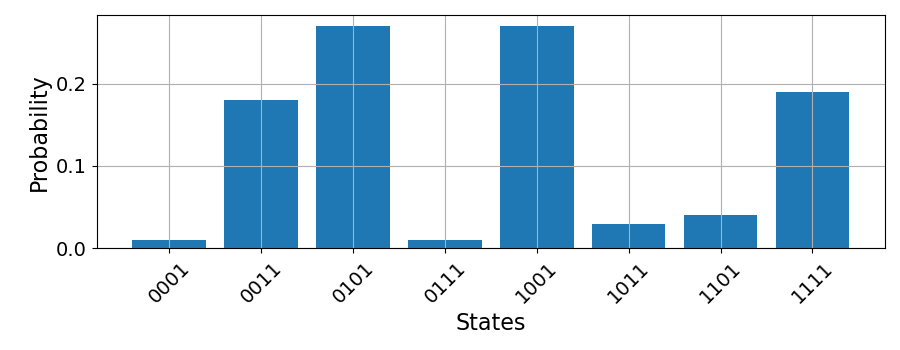}
\caption{Simulation results of 100 runs for a single tile of an LHZ network implemented with CMOS. The figure shows that the probability of states with an even number of '1' spins is high and that the differences in the probability of such even states are small.}
\vspace{-1em}
\label{fig:SimResultCMOS}
\end{figure}

\begin{figure}[ht]
\centering
\begin{subfigure}[b]{.49\linewidth}
    \centering
    \begin{tikzpicture}
        \node[anchor=south west,inner sep=0] (image) at (0,0)
        {\includegraphics[width=.99\linewidth]{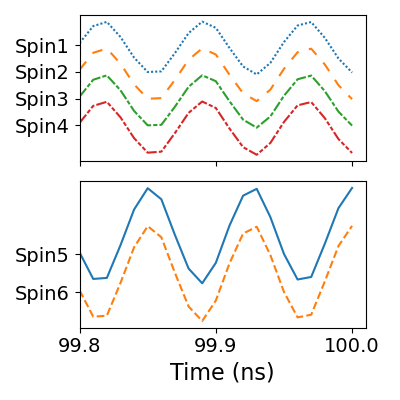}};
        \draw (image.north west) node[right] {(a)};
    \end{tikzpicture}
\end{subfigure}
\begin{subfigure}[b]{.49\linewidth}
    \centering
    \begin{tikzpicture}
        \node[anchor=south west,inner sep=0] (image) at (0,0)
        {\includegraphics[width=.99\linewidth]{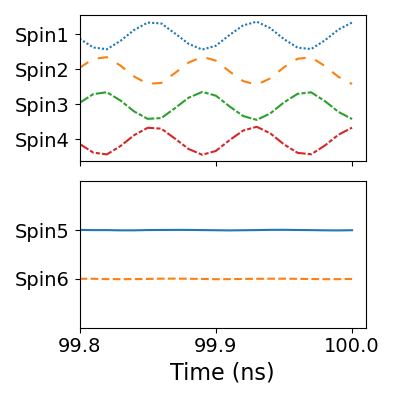}};
        \draw (image.north west) node[right] {(b)};
    \end{tikzpicture}
\end{subfigure}

\caption{Simulation results for each node normalized to the maximum amplitude of the Spin 1 is demonstrated for two different outcomes: (a) four LHZ nodes (Spin 1 to 4) are in-phase, and the two internal spins of the ancillary node (Spin 5 and 6) are out of phase with the nodes ensuring minimum energy; (b) two LHZ nodes (Spin 1 and 3) are in-phase, and two out of phase (Spin 2 and 4), and the amplitudes of the two internal spins of the ancillary node (Spin 5 and 6) are negligible, resulting in ground state. These results show an even number of ones we expect from a tile for LHZ implementation.}
\vspace{-1em}
\label{fig:WaveformCMOS}
\end{figure}

The proposed analog LHZ tile has been implemented in Cadence Virtuoso with 12nm technology and simulated using Cadence Spectre. The simulation result of 100 runs for a single LHZ tile is shown in Fig.~\ref{fig:SimResultCMOS}. Here, the probability of correct states is around \qty{90}{\percent}, and the probability distribution is relatively uniform for each possible correct state; the two facts indicate that the tile circuit is suitable for an LHZ IM. This design takes about \qty{31}{\ns} to settle in the final state. The transients depicted in Fig.~\ref{fig:WaveformCMOS} are the voltage waveforms at the four physical spins (i.e., Spin 1 to 4) and two internal spins of the ancillary node (Spin 5 and 6). The waveforms are normalized to the Spin 1's amplitude. Fig.~\ref{fig:WaveformCMOS}~(a) shows one possible minimum-energy state: all of the four nodes in an LHZ tile are in-phase and out of phase with the two ancillary nodes. Fig.~\ref{fig:WaveformCMOS}~(b) shows another least-energy configuration where two LHZ spins in-phase and two out of phase, and the sum of ancillary nodes' energy near zero, resulting in a ground state.

\begin{table}[htbp]
    \centering
    \caption{Fully connected 100-node state-of-the-art Ising machines}
    \resizebox{\columnwidth}{!}{%
    \begin{tabular}{c|c|c|c|c}
        \hline
         & \textbf{\cite{moy2022ringOsc}} & \textbf{\cite{wang2024capCoupledLatch}} & \textbf{\cite{wang2024capCoupledLatch}} & \textbf{This work} \\
         \hline\hline
         Architecture & King's graph & Cross-bar & Cross-bar & LHZ \\
         \hline
         Coupling & \makecell{Transmission\\ gates} & Resistive & Capacitive & Resistive \\
         \hline
         \makecell{Time-to-solution} & \qty{30}{\us} & \qty{0.94}{\ns} & \qty{64.64}{\ns} & \qty{100}{\ns} \\
         \hline
         \makecell{Energy per\\ inference}& \qty{76.8}{\nano\J} & \qty{2.12}{\mJ} & \qty{4.85}{\uJ} & \qty{2.91}{\uJ} \\
         \hline
         \makecell{Energy-time\\ product (\unit{\J\cdot\s})} & \num{2.30e-12} & \num{1.99e-12} & \num{3.14e-13} & \num{2.91e-13} \\ 
         \hline
         \makecell{Average power\\ (\unit{\W})} & \num{2.56e-3} & \num{2.26e6} & 75.0 & 29.1 \\ \hline
    \end{tabular}
    }\label{Tab:Comparison}
\end{table}

IM efficiently solves QUBO problems with faster response times and lower energy consumption than classical machines. Using the example in \cite{wang2024capCoupledLatch} of a 100-node MaxCut graph with a degree of three, we calculate power consumption based on the required tiles. Simulated results in~\cite{wang2021timeScaling} suggest that the time to reach a low-energy state does not scale significantly with increased problem size. Our analysis estimates the inference time for a 100-node MaxCut problem is about~\qty{100}{\ns} and consumes approximately \qty{2.91}{\uJ}. This estimation yields an energy-time product of \qty{2.91e-13}{\J\cdot\s}, comparable to the best result in \cite{wang2024capCoupledLatch} as seen in Table~\ref{Tab:Comparison}. To reduce circuit power, we propose powering only a few non-ancillary nodes in the LHZ architecture while keeping ancillary nodes active and decreasing bias currents to lower oscillating amplitudes, significantly improving energy per problem.

It should be noted that the main factors in energy consumption are power and time-to-solution. Digital annealing processors feature low power consumption but take much longer to settle in the ground state. For example, the design~\cite{Megumi2024CMOS22N} consumes \qty{35.2}{\mW} at \qty{10}{\MHz} yet it takes 10,000 steps to solve a 512-node Max-cut problem. On the other hand, analog implementations solve such problems significantly faster but with a higher power consumption. For instance, \cite{wang2024capCoupledLatch} report a 100-node Max-cut solver with \qty{64.64}{\ns} time-to-solution and \qty{75}{\W} average power consumption. Our proposed method is also an analog implementation and has a high power consumption of 5mW per tile. However, the fast time to solution makes it more energy efficient compared to other designs.

\section{Conclusion}
The LHZ-based Ising machine promises an efficient solution for NP-complete problems with a scalable architecture. We have proposed the basic element (i.e., tile) of an analog LHZ Ising machine with an acceptable accuracy that can be utilized in a conventional CMOS process. To guarantee the correct functionality of the LHZ tile, we suggested a circuit implementation for the ancillary node. We implemented the proposed analog tile in 12nm technology and verified its functionality and stability. The tile stabilizes at 31ns and consumes \qty{5}{\mW}.

\vfill

\end{document}